%% file: main.tex
\documentclass[review]{elsarticle}

\usepackage{float}
\usepackage{subcaption}
\usepackage{relsize}
\usepackage{etoolbox,enumitem}
\usepackage{amssymb,amsmath,amsthm,enumitem}
\usepackage[ruled,vlined]{algorithm2e}
\usepackage{ mathrsfs }
\usepackage[framed,numbered,autolinebreaks,useliterate]{mcode}
\usepackage{listings}
\usepackage{showexpl}
\usepackage[most]{tcolorbox}
\usepackage[colorlinks = true,
            linkcolor = black,
            urlcolor  = blue,
            citecolor = black,
            anchorcolor = black]{hyperref}

\newcommand{\code}[1]{\texttt{#1}}

\newcommand{\newpar}{{}}
\newcommand{\acf}{{\text{ACF}}}
\newcommand{\iac}{{\text{IAC}}}

\newcommand{\Pam}{{Q}} 
\newcommand{\pam}{{q}} 
\newcommand{\diff}{{\operatorname{d}}}
\newcommand{\cov}{{\text{COV}}}

\newcommand{\chainSize}{{N_{\mathrm{sample}}}}
\DeclareMathOperator*{\argmax}{\operatorname*{arg\,max}}

\journal{Journal of Computational Physics}

\begin{document}

\begin{frontmatter}

\title{MatDRAM: A pure-MATLAB Delayed-Rejection Adaptive Metropolis-Hastings Markov Chain Monte Carlo Sampler}


\author{Shashank Kumbhare\fnref{SK}, Amir Shahmoradi\fnref{AS}}
\address{Department of Physics, The University of Texas, Arlington, TX, United States}
\fntext[SK]{shashank.kumbhare@mavs.uta.edu}
\fntext[AS]{a.shahmoradi@uta.edu}




\begin{abstract}
Markov Chain Monte Carlo (MCMC) algorithms are widely used for stochastic optimization, sampling, and integration of mathematical objective functions, in particular, in the context of Bayesian inverse problems and parameter estimation. For decades, the algorithm of choice in MCMC simulations has been the Metropolis-Hastings (MH) algorithm. An advancement over the traditional MH-MCMC sampler is the Delayed-Rejection Adaptive Metropolis (DRAM). In this paper, we present MatDRAM, a stochastic optimization, sampling, and Monte Carlo integration toolbox in MATLAB which implements a variant of the DRAM algorithm for exploring the mathematical objective functions of arbitrary-dimensions, in particular, the posterior distributions of Bayesian models in data science, Machine Learning, and scientific inference. The design goals of MatDRAM include nearly-full automation of MCMC simulations, user-friendliness, fully-deterministic reproducibility, and the restart functionality of simulations. We also discuss the implementation details of a technique to automatically monitor and ensure the diminishing adaptation of the proposal distribution of the DRAM algorithm and a method of efficiently storing the resulting simulated Markov chains. The MatDRAM library is open-source, MIT-licensed, and permanently located and maintained as part of the ParaMonte library at: \url{https://github.com/cdslaborg/paramonte}.
\end{abstract}


\end{frontmatter}


\setlength{\abovedisplayskip}{3pt}
\setlength{\belowdisplayskip}{3pt}
\flushbottom
\maketitle
\thispagestyle{empty}

\input{sec/intro}

\input{sec/methods}

\input{sec/results}
\input{sec/discussion}



\bibliography{../../../libtex/all}


\end{document}

%% file: sec/intro.tex
\section{Introduction}
\label{sec:intro}

    At the foundation of predictive science lies the scientific methodology, which involves multiple steps of observational data collection, developing testable hypotheses, and making predictions. Once a scientific theory is developed, it can be cast into a mathematical model whose parameters have to be fit via observational data. This leads to the formulation of a mathematical objective function for the problem at hand, which has to be then optimized to find the best-fit parameters of the model or sampled to quantify the uncertainties associated with the parameters, or integrated to assess the performance of the model.

\begin{figure}
        \centering
        \fbox{
            \begin{subfigure}{.49\linewidth}
                \centering
                \includegraphics[width=.98\linewidth]{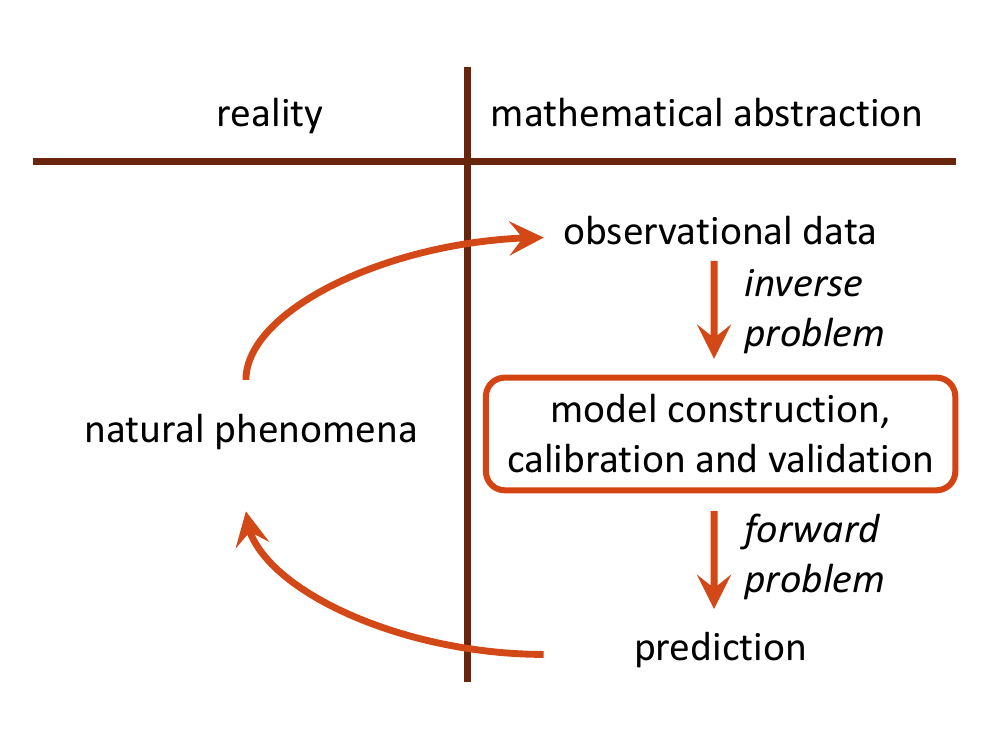}
                \caption{The scientific methodology}
                \label{fig:scientific methodology}
            \end{subfigure}
            \hfill                                                    
            \begin{subfigure}{.49\linewidth}
                \centering
                \includegraphics[width=.98\linewidth]{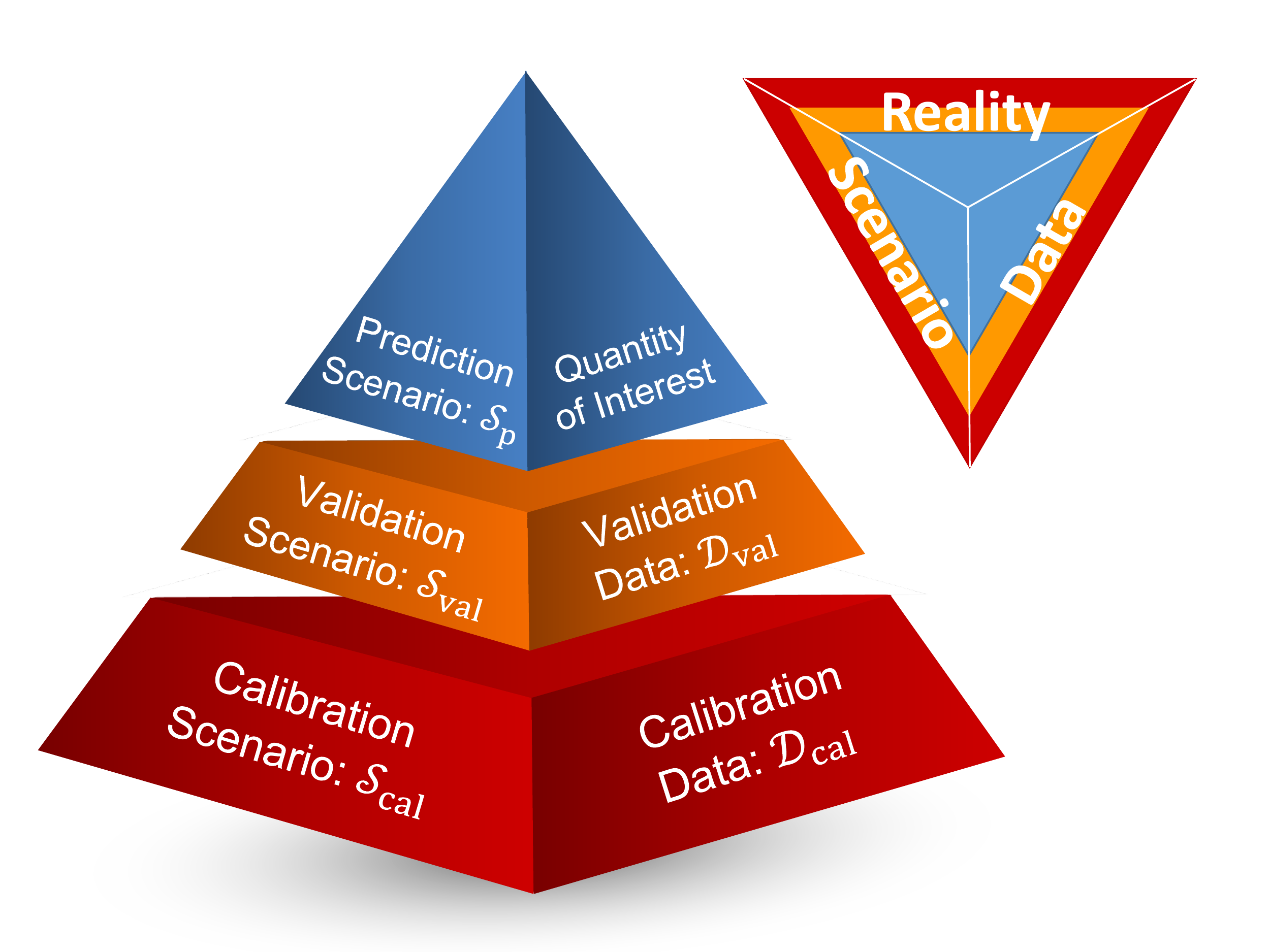}
                \caption{The prediction pyramid}
                \label{fig:prediction pyramid}
            \end{subfigure}
        }
        \caption{\textbf{(a) An illustration of the scientific methodology:} The left hand side of table represents the real world in which natural phenomena occur. The right hand side represents the mathematical abstraction of reality. First, an initial experiment is performed and observational data about a natural phenomenon is collected. Then, a hypothesis is formed based upon which a mathematical model is constructed. The model is subsequently calibrated and validated, and finally, predictions about the quantities of interest are made, whose accuracies can be tested against further independent observational data. \textbf{(b) The prediction pyramid}, depicting the three hierarchical levels of predictive inference from bottom to top: Calibration, Validation, and Prediction of the Quantity of Interest. The rear face of the tetrahedron represents the reality (truth) about the set of observed phenomena, which is never known to the observer. The right-front face of the tetrahedron represents the observational data, which is the truth convolved with various forms of uncertainties. The left-front face represents the scenarios, under which data is collected, as well as the set of models that are hypothesized to describe the unknown truth \cite{shahmoradi2017multilevel, 2017arXiv171110599S, 2020arXiv200809589S, shahmoradi2020paradram}.}
        \label{fig:scientific inference}
    \end{figure}

    The process of {\it parameter tuning} is also commonly known as the \textit{model calibration} during which the free parameters are constrained. Subsequently, to demonstrate the effectiveness, accuracy and reliability of the model, it has to go through the \textit{validation} step, where the performance of model is tested against a dataset independently of the calibration or training data. Once validated, the model can be used to predict the \textit{Quantity of interest} (QoI) in the inference problem. This workflow is illustrated in Figure \ref{fig:scientific methodology}, which can be also represented more formally in a hierarchical pyramid structure known as the \textit{prediction pyramid} as shown in Figure \ref{fig:prediction pyramid}.

    \subsection{The optimization and sampling of the objective function}

\begin{figure}
            \centering
            \fbox{\includegraphics[width=0.49\linewidth]{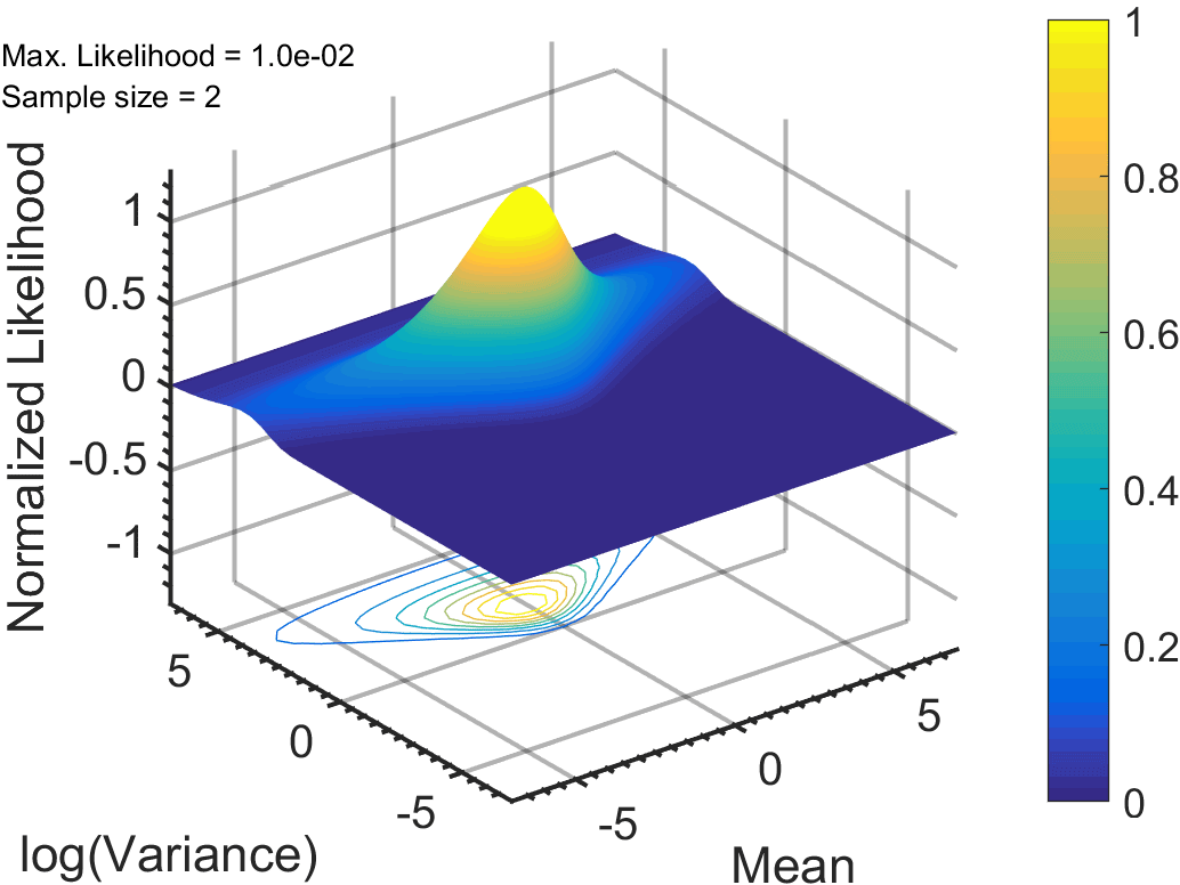}}
            \caption{\textbf{An illustration of the mathematical objective function}: The color-coded surface plot represents the likelihood function of a dataset composed of only two single-valued observations, under the hypothesis of having been generated from a Gaussian distribution. The domain of the objective function is comprised of the parameters of the Gaussian distribution (e.g., mean and variance) whose best-fit values and their associated uncertainties must be evaluated.}
            \label{fig:Objective Function}
        \end{figure}

        The scientific prediction problems generally require the formulation of a mathematical objective function whose value represents the goodness of a possible solution to the problem. Among the most popular choices of such objective functions is the posterior probability density (PPD) of the parameters of a model, or in particular scenarios, the likelihood function. In such cases, the process of inference involves the maximization of the PPD or the likelihood function to find the optimal solution to the problem among potentially infinite number of solutions. An example likelihood function for a very simple problem of inferring the parameters of a Gaussian Probability Density Function (PDF) given only two observational data points is illustrated in Figure \ref{fig:Objective Function}. The set of parameters for which the value of objective function is a global maximum (or in come cases, minimum) is commonly called the \textit{best-fit} parameters.

        Ever since the invention of digital computers, deterministic and stochastic optimization techniques have become indispensable tools for inferring the best-fit parameters of scientific models. These algorithms, however, inherently lack the ability to explore the entire domain of the objective function. Such exploration is essential to form confidence bounds on the best-fit solutions to a problem and to quantify the uncertainty in the inference. For decades, full exploration of the complex mathematical objective functions were nearly impossible. The advent of new computational technologies over the past few decades, however, has dramatically changed the landscape of scientific inference and has led to the emergence new fields of science, such as {\it Uncertainty Quantification} and has led to their exponential growth over the past years \citep{shahmoradi2017multilevel}. Consequently, sampling techniques, in particular, Monte Carlo methods \citep{metropolis1949monte, von195113, segre1955fermi, metropolis1987beginning} have become routine tools in scientific inference problems and uncertainty quantification.

        Among the most successful and popular Monte Carlo techniques is the Markov Chain Monte Carlo (MCMC), first introduced and discussed by \cite{metropolis1953equation} for solving numerical Physics problems. The Metropolis Algorithm of \cite{metropolis1953equation} was subsequently generalized by \cite{hastings1970monte} and further popularized in the Statistics community as what has now become known as the Metropolis-Hastings MCMC algorithm for sampling of mathematical objective functions.

        Optimization and Monte Carlo techniques have played a major role in the advancement of science over the second half of the twentieth century as well as the new millennium. Given their fundamental importance and mathematically-sound foundations, their popularity will only grow in the future. In particular, the MCMC techniques have become popular practical tools in many fields of science and engineering, from Astrophysics \citep[e.g., ][]{shahmoradi2013multivariate,shahmoradi2015short, 2020arXiv200601157O, 2019arXiv190306989S} to Bioinformatics and Biomedical Sciences \citep[e.g., ][]{shahmoradi2014predicting, lima2017ices, lima2017selection}.

        Despite the popularity of the MCMC technique, this method, as presented in its original form by \cite{metropolis1953equation} and \cite{hastings1970monte}, has a significant drawback. The traditional MCMC methods often require hand-tuning of several parameters within the sampling algorithms to ensure fast convergence of the resulting Markov chain to their target densities for a particular problem at hand. Countless studies have been published, in particular during the last decade of the twentieth century along with the rise of personal computers, to bring full automation to the problem of tuning of the free parameters of Markov Chain Monte Carlo samplers. Among the most successful is the algorithm of \citep{haario2006dram}, famously known as the {\bf D}elayed-{\bf R}ejection {\bf A}daptive {\bf M}etropolis MCMC or {\bf DRAM}.

    \subsection{Existing computational toolboxes}

        Several computational toolboxes in different programming languages currently implement variants of DRAM or the traditional MCMC algorithms. Open-source examples include \code{FME} \citep{soetaert2010inverse} in R, \code{PyMC} \citep{patil2010pymc} and \code{pymcmcstat} \citep{miles2019pymcmcstat} in Python, \code{mcmcstat} \citep{haario2006dram} and PESTO \citep{stapor2018pesto} in MATLAB, \code{mcmcf90} \citep{haario2006dram} in Fortran, and QUESO \citep{queso} in C/C++ programming languages. To the best of our knowledge, these packages are bound to a single high-level programming language or small family of programming languages, and some (e.g., QUESO) heavily rely on external library dependencies that require significant care, maintenance, and time, up to a full day, for a proper build and installation. In the case of MATLAB programming environment, very few options exist for the MCMC/DRAM sampling problems and among available options, not all the desired functionalities are implemented yet, including comprehensive reporting, automatic restart functionality, and nearly-full automation of simulations. 
        \newpar

        Most recently, we have attempted to circumvent some of the aforementioned limitations of the existing MCMC packages by developing the ParaMonte library \cite{shahmoradi2019paramonte, kumbhare2020parallel, 2020arXiv200809589S, shahmoradi2020paradram, shahmoradi2020paramonte, 2020arXiv200914229S, shahmoradi2020paramonteII, 2020arXiv201000724S, kumbhare2020parallel}, a cross-language toolbox for serial and parallel Monte Carlo simulations, accessible to a wide range of programming languages.

    \subsection{The ParaMonte MatDRAM library}

        To address the aforementioned heterogeneities and shortcomings in the existing implementations of the DRAM algorithm, \cite{2020arXiv200809589S, 2020arXiv201000724S, 2020arXiv200914229S} have recently implemented the ParaMonte library, which provides a unified Application Programming Interface and environment for serial and parallel MCMC, DRAM, and other Monte Carlo simulation techniques accessible from a wide range of programming languages, including C, C++, Fortran, and Python. In particular, the ParaDRAM algorithm within the ParaMonte library provides a high-performance, fully-automated parallel implementation of a variant of the DRAM algorithm of \citep{haario2006dram} that is directly accessible from within the popular compiled programming languages (C/C++/Fortran). The ParaMonte library has been designed while bearing the following design philosophy and goals in mind,

        \begin{enumerate}[parsep=0ex,itemsep=1ex,topsep=1ex]
            \item {\bf Full automation} of all Monte Carlo simulations to ensure the highest level of user-friendliness of the library and minimal time investment requirements for building, running, and post-processing of MCMC simulations.
            \item {\bf Interoperability} of the core library with as many programming languages as currently possible.
            \item {\bf High-Performance} meticulously-low-level implementation of the library to ensure the fastest-possible Monte Carlo simulations.
            \item {\bf Parallelizability} of all simulations via two-sided and one-sided MPI/Coarray communications while requiring zero-parallel-coding efforts by the user.
            \item {\bf Zero-dependence} on external libraries to ensure hassle-free Monte Carlo simulation builds and runs.
            \item {\bf Fully-deterministic reproducibility} and automatically-enabled restart functionality for all simulations up to 16 digits of precision if requested by the user.
            \item {\bf Comprehensive-reporting and post-processing} of each simulation and its results, as well as their automatic compact storage in external files to ensure the simulation results will be comprehensible and reproducible at any time in the distant future.
        \end{enumerate}

        Following the design goals of the ParaMonte library, here we present the MatDRAM library, which is a pure-MATLAB implementation of the ParaDRAM reference algorithm in the ParaMonte library. MatDRAM can be considered as an extension of the ParaDRAM algorithm to the MATLAB programming environment, albeit currently without the parallelism features of ParaDRAM. The MatDRAM library that we present in this manuscript has been carefully designed to mimic and reproduce virtually all of the functionalities and characteristics of the serial version of the ParaDRAM algorithm in C/C++/Fortran, including the restart functionality. This feature is very useful for situations where runtime interruptions happen at any stage during the simulation. Despite the stochastic nature of Monte Carlo simulations, the MatDRAM sampler is able to resume an interrupted simulation from where it left off and reproduce {\it the same results}, up to 16 digits of precision, that it would have produced without the interruption.

        MatDRAM additionally has the post-processing tools that enable seamless analysis and visualization of the simulation results, whether performed via MatDRAM, or ParaDRAM, regardless of the programming language environment (Python/MATLAB/Fortran/C++/C) or the platform (Windows/Linux/macOS) used for the simulation. These post-processing tools have been deliberately designed to highly resemble the post-processing tools of the \code{ParaMonte::Python} library \citep{2020arXiv201000724S}.

        In the following sections, we present algorithmic and implementation details of the MatDRAM library, including a discussion of the specific variant of the DRAM algorithm that we have implemented in \S\ref{sec:methods}, as well as the methodology that we introduce to automatically monitor and ensure the diminishing adaptation of the proposal distribution of the DRAM algorithm. We also discuss an efficient method of storing the resulting MCMC chains from MatDRAM which can reduce the output overhead and memory requirements of the algorithm, on average, by a factor of 4 to 10 or more, depending on the scale and complexity of the simulation. We discuss the Application Programming Interface of MatDRAM and example simulations in \S\ref{sec:results}, followed by discussion and concluding remarks in \S\ref{sec:discussion}.

%% file: sec/methods.tex
\section{Methodology}
\label{sec:methods}

    The MatDRAM library is based on the algorithm of Delayed-Rejection Adaptive Metropolis-Hastings (DRAM) algorithm of \citep{haario2006dram}, which can be regarded as an extension of the traditional the Metropolis-Hastings (MH) algorithm. The MH algorithm first appeared in \cite{metropolis1953equation} and was later generalized by \cite{hastings1970monte} to a broader set of stochastic samplers. The DRAM algorithm itself is a combination of two powerful ideas that have appeared in the MCMC literature: Adaptive Metropolis (AM) \citep{haario1999adaptive,haario2001adaptive}, and Delayed Rejection (DR) \citep{tierney1999some, green2001delayed, mira2001metropolis}. In the following sections, we briefly introduce the standard MH algorithm, the AM algorithm, the DR algorithm, the combination of DR and AM (i.e., DRAM) algorithms, and finally, the specific variant of the DRAM algorithm that we have implemented in the MatDRAM library.

    \subsection{The Metropolis-Hastings (MH) algorithm}

        The Metropolis-Hastings (MH) algorithm is the most popular MCMC method and has been widely adopted by the community. In MH algorithm, a Markov chain is constructed as a progressively more-detailed picture of the target distribution, starting from some point within the domain of the target objective function.

        Let $\pi$ be the target distribution, which we wish to sample from, defined on some finite discrete state space $\chi$. The MH algorithm, named after \cite{metropolis1953equation} and \citep{hastings1970monte}, proposes a mechanism to construct a Markov sequence of random variables $X_1,X_2,...$ on $\chi$ such that the resulting chain is ergodic and stationary with respect to $\pi$, that is, if $X_t$ $\sim$ $\pi$(x), then $X_{t+1}$ $\sim$ $\pi(x)$ and therefore, the sequence converges in distribution to $\pi$.

        Let $X_0$ be the starting point of our Markov Chain which we can choose arbitrarily and $X_t=x$ be the current state at step $t$. The MH algorithm associated with the target density $\pi$ requires a conditional density $q$ called the {\it proposal kernel} or {\it proposal distribution}. The transition from $X_t$ at time $t$ to $X_{t+1}$ is done via the following steps:

        \begin{enumerate}[parsep=0ex,itemsep=1ex,topsep=1ex]
            \item Generate $Y_t\sim q(\cdot\mid x)$
            \item Accept $Y_t$ with the probability $\alpha$ and set,
                \begin{equation}
                    X_{t+1} = 
                    \begin{cases}    
                        Y_{t},\quad   \text{with probablility} \quad \alpha(x,y) \\    
                        x,\quad   \text{with probablility} \quad 1\ -\ \alpha(x,y)    
                    \end{cases}
                \end{equation}
                \noindent where,
                \begin{equation}
                    \label{accProb}
                    \alpha(x,y) = 1 \wedge     \dfrac{\pi(y)}{\pi(x)} ~ \dfrac{q(y \mid x)}{q(x \mid y)}
                \end{equation}
        \end{enumerate}

    \subsection{The Adaptive Metropolis (AM) algorithm}

        The Adaptive Metropolis (AM) strategy is discussed in detailed in \cite{haario2001adaptive}. The distinctive idea in the AM as opposed to the MH algorithm is to create a Gaussian proposal kernel $q$ with a covariance matrix calibrated using the previously-sampled path of the MCMC chain. Therefore, the covariance matrix of the proposal distribution $q$ of the AM algorithm, is not only dynamically varying, but this variation also depends on the history of the chain, i.e., the points sampled so far. After an initial non-adaptation period, say $t_0$, $q$ will be centered at the last sampled state, $X_t$, with covariance $C_t$ = $s_d.Cov(X_0,...,X_{t-1})+s_d\varepsilon I_{d}$, where $s_d$ is a parameter that depends only on dimension $d$ of $\chi$ on which $\pi$ is defined, $\varepsilon > 0$ is a constant that we may choose to be very small and positive to ensure the positive-definiteness of the covariance matrix, and $I_d$ denotes the $d$-dimensional identity matrix. At the start of the MCMC simulation, an arbitrarily-defined positive definite covariance matrix, $C_0$, is specified by the user. This initial covariance matrix will hopefully represent our closest guess for the shape and scale of the covariance matrix of the target density function. Then, the covariance matrix of the proposal distribution of the MCMC sampler at any stage during the simulation is defined as,
        \begin{equation}
            C_t = \begin{cases}    C_0,\qquad\qquad\qquad\qquad\qquad\quad t\leq t_0   \\            
                                s_d ~ \cov(X_0,...,X_{t-1})+s_d \varepsilon I_{d}, \quad t>t_0        
                    \end{cases}
            \label{covAdapted}
        \end{equation}
        where,
        \begin{equation}
            \cov(X_0,...,X_k) = \dfrac{1}{k}                                                                                    
                                \left( \mathlarger{\sum}_{i=0}^{k} X_iX^T_i-(k+1)\overline{X_k}\; \overline{X}_k^T \right)     
        \end{equation}
        where $\overline{X_k} = \tfrac{1}{k+1} \sum_{i=0}^k X_i$

        In general, the adaptation does not necessarily need to be performed at every MCMC step, but rather for certain time intervals. This form of adaptation improves the mixing properties of the algorithm. In this context, the index $t_0$ could be in fact used to define the length of non-adaptation periods during the MCMC sampling.

        Based on the findings of \cite{gelman1996efficient}, the scaling parameter is typically set to $s_d = 2.4^2/d$ where $d$ represents the number of dimensions of the domain of the objective function. This specific value has been shown to optimize the mixing properties of the MCMC sampler in the case of an infinite dimensional Standard MultiVariate Normal (MVN) distribution when explored by an MVN proposal distribution.

    \subsection{The Delayed Rejection (DR) algorithm}

        In the traditional MH algorithm, a new state is drawn from the proposal distribution $q(\cdot|\cdot)$ at each stage of the MCMC sampling, which is subsequently either accepted or rejected with the probability given by Equation \eqref{accProb}. If rejected, the chain remains in the current state and a new proposal is drawn. This approach is quite useful when the domain of the objective function is low-dimensional. In the case of high-dimensional domains, hoever, the number of rejections can become significantly larger than the acceptances, leading to a dramatic decrease in the efficiency of the sampler.

        To overcome this problem of high rejection rate, a strategy called Delayed Rejection (DR) was proposed by \cite{green2001delayed}. Reducing the number of rejected proposals is one of the most important goals in every MCMC application. By doing this, one improves the MCMC sampling efficiency in the Peskun (1973) sense \citep{peskun1973optimum}. In DR, when a new state is proposed and rejected, another state is proposed with a different distribution which may depend on the rejected state. Then, the newly proposed state is either accepted or rejected with a suitable acceptance probability that preserves the detailed balance of the MCMC chain. If the proposed state is rejected, a third state can be proposed. Since the entire process is time-reversible, this process of delaying the rejection can continue for as long as desired. This method is particularly useful for multimodal target density functions whose modes are separated by deep valleys of low likelihoods.

        The generic procedure to obtain the $i$th-stage delayed-rejection proposal is as follows (the superscript denotes the DR stage):

        \begin{enumerate}[parsep=0ex,itemsep=1ex,topsep=1ex]

            \item Generate $Y^0_t\sim q^0(\cdot\mid x)$

            \item Accept $Y^0_t$ with probability $\alpha^0$
                \begin{equation}
                    X_{t+1} = 
                    \begin{cases}
                        Y^0_{t},\quad   \text{with probability}: \; \alpha^0(x,y^0) \\    
                        x,\quad \;\;\;  \text{with probability}: \; 1 - \alpha^0(x,y^0)     
                    \end{cases}
                \end{equation}
                \noindent where,
                \begin{equation}
                    \label{accProbDR0}
                    \alpha^0(x,y^0) = 1 \wedge     \dfrac{\pi(y^0)}{\pi(x)} ~ \dfrac{q^0(y^0 \mid x)}{q^0(x \mid y^0)}
                \end{equation}

            \item If $Y^0_t$ is rejected, \\
                $1^{st}$ DR stage : Generate $Y^1_t\sim q^1(\cdot\mid y^0,x)$

            \item Accept $Y^1_t$ with probability $\alpha^1$ \\ 
                where,
                \begin{eqnarray}
                    \label{accProbDR1}
                    \alpha^1(x,y^0,y^1) &=& 1 \wedge    \dfrac{\pi(y^1)}{\pi(x)}                             
                                                        \dfrac{q^0(y^1 \mid y^0)}{q^0(x \mid y^0)} ~            
                                                        \dfrac{q^1(y^1 \mid y^0,x)}{q^1(x \mid y^0,y^1)} ~    
                                                        \dfrac{[1-\alpha^0(y^1,y^0)]}{[1-\alpha^0(x,y^0)]} \\    
                                        &\vdots& \nonumber
                \end{eqnarray}
            \item If $Y^{i-1}_t$ is rejected, \\
                $i^{th}$ DR stage : Generate $Y^i_t\sim q^i(\cdot\mid y^{i-1},...,y^0,x)$
            \item Accept $Y^i_t$ with probability $\alpha^i$ \\
                where,
                \begin{align}
                    \alpha^i(x,y^0,...,y^i) & = 1 \wedge     \bigg\{
                                \dfrac{\pi(y^i)}{\pi(x)}~                                                    
                                \dfrac{q^0(y^i \mid y^{i-1})}{q^0(x \mid y^0)}~                                
                                \dfrac{q^1(y^i \mid y^{i-1},y^{i-2})}{q^1(x \mid y^0,y^1)}\ldots                
                                \dfrac{q^i(y^i \mid y^{i-1},\ldots,y^0,x)}{q^i(x \mid y^0,\ldots,y^i)} \nonumber\\
                            &     \dfrac{[1-\alpha^0(y^i,y^{i-1})]}{[1-\alpha^0(x,y^0)]}~                        
                                \dfrac{[1-\alpha^1(y^i,y^{i-1},y^{i-2})]}{[1-\alpha^1(x,y^0,y^1)]}\ldots        
                                \dfrac{[1-\alpha^{i-1}(y^i,\ldots,y^0)]}{[1-\alpha^{i-1}(x,y^0,\ldots,y^{i-1})]}    
                                                            \bigg\}
                \label{accProbDRi}
                \end{align}

        \end{enumerate}

    \subsection{The Delayed-Rejection Adaptive-Metropolis (DRAM) algorithm}

        The performance of the MCMC sampler can be even further improved by combining the Delayed Rejection (DR) algorithm with the Adaptive Metropolis (AM) algorithm, resulting in the DRAM algorithm, as described in detail by \cite{haario2006dram}. While the AM algorithm adapts the proposal distribution $q(\cdot|\cdot)$ based on the past history of the chain, the DR algorithm improves the efficiency of the resulting MCMC estimator. In other words, AM allows for {\it global} adaptation of the proposal distribution based on all previously accepted proposals, while DR allows for {\it local} adaptation, only based on rejected proposals within each time-step.

        Given an initial non-adaptation period of length $t_0$ and an initial covariance matrix for the proposal distribution $C_0$, the DRAM algorithm can be described for any time step $t$ as the following,

        \begin{enumerate}[parsep=0ex,itemsep=1ex,topsep=1ex]

            \item Update the covariance matrix $C^0_t$ for the proposal kernel $q^0$:
                \begin{equation}
                    C^0_t=\begin{cases}    C_0,\qquad\qquad\qquad\qquad\qquad\quad \:                t\leq t_0        \\ 
                                        s_d~\cov(X_0,...,X_{t-1})+s_d \varepsilon I_{d}, \quad   \text{mod}(t,t_0) = 0     \\ 
                                        C^0_{t-1}, \qquad\qquad\qquad\qquad\qquad \;            \text{otherwise}          \\ 
                            \end{cases}
                    \label{covAdaptedDRAM}
                \end{equation}

            \item Generate $Y^0_t\sim q^0(\cdot\mid x)$ and accept with probability $\alpha^0$ given by Equation \eqref{accProbDR0}.

            \item For the $i^{th}$ DR stage ($i=1,2,3...,m$):
                \begin{enumerate}[parsep=0ex,itemsep=0.6ex,topsep=0.5ex]
                    \item If $Y^{i-1}_t$ is rejected,
                    \item $q^{i}(\cdot\mid y^{i-1},\ldots,y^0,x) = \gamma^i \times q^{i-1}(\cdot\mid y^{i-2},\ldots,y^0,x) $ 
                    \item Generate $Y^i_t\sim q^{i}(\cdot\mid y^{i-1},\ldots,y^0,x)$
                    \item Accept with probability $\alpha^i$ given by Equation \eqref{accProbDRi}.
                \end{enumerate}

        \end{enumerate}

        The scale factor $\gamma^i$ can be freely chosen. The simulation results in \cite{green2001delayed} suggest that it is more beneficial, in terms of asymptotic variance reduction of the resulting estimators, to have larger scale factors at earlier stages which become progressively smaller upon each delayed rejection. The advantage of such scaling might be already clear from the above discussions: If the initial guess for the shape and scale of the proposal distribution $q(\cdot|\cdot)$ is far from the optimal shape and scale, it may be difficult to get the adaptation process started. This happens if the variance of $q(\cdot|\cdot)$ is too large, or if the covariance for the proposal is nearly singular. In either case, no proposed states are practically accepted. The remedy for this problem is to reduce the variance of $q^i$ in the higher stages of DR, increasing the probability of points to be accepted.

    \subsection{The MatDRAM algorithm}

        One of the major weaknesses of the delayed-rejection algorithm is that a strategy is generally required to construct a set of delayed-rejection-stage proposal distributions that utilize the information collected from the rejected states during the DR process. Moreover, the equation for the acceptance probability during the DR process becomes progressively more complex with increasing the number of delayed rejection stages.

        The two aforementioned challenges make a practical implementation of the DR algorithm nearly impossible. However, the implementation of the DR process can be greatly simplified if we limit our attention to symmetric proposal distributions whose shape remain fixed throughout the DR process and whose scales are determined by a scaling schedule pre-specified by the user prior to the simulation. The MatDRAM algorithm that we present in this work implements this specific variant of the generic DRAM algorithm of \citep{haario2006dram}.

        This symmetric Delayed-Rejection sampling scheme has been discussed by \cite{mira2001metropolis} where the proposal kernel is symmetric and is allowed to depend only on the last rejected state,
        \begin{equation}
            q(y_i \mid x, \{y\})= q(y_i \mid y_{i-1}), \qquad i = 1, 2, \ldots, N
            \label{symKernelMat}
        \end{equation}
        \noindent where, $\{y\}$ is the set of $N$ states that have been visited and rejected so far.

        In this setting, Equation \eqref{accProbDR0} reduces to,
        \begin{equation}
            \alpha^0(x,y^0) = 1 \wedge \dfrac{\pi(y^0)}{\pi(x)}
            \label{accProbDR0Mat}
        \end{equation}
        \noindent for the $0^{th}$ DR stage. Combining Equation \eqref{symKernelMat} and Equation \eqref{accProbDR0Mat} with Equation \eqref{accProbDR1}, the acceptance probability for first DR stage becomes,
        \begin{equation}
            \alpha^1(x,y^0,y^1) = 1 \wedge    \dfrac  { \pi(y^1) \left[1 - 1 \wedge \dfrac{\pi(y^0)}{\pi(y^1)} \right] }     
                                                    { \pi(x)-\pi(y^0) }                                                 
            \label{accProbDR1Mat}
        \end{equation}

        Three cases can occur at this point:
        \begin{enumerate}[parsep=0ex,itemsep=1ex,topsep=1ex]
            \item if $\pi(y^1)         \geq    \pi(x)      $ then $\alpha^1(x,y^0,y^1) = 1$, thus accept $y^2$ and set $X_{t+1}=y^1$.
            \item if $\pi(y^1)         <        \pi(y^0) $ then $\alpha^1(x,y^0,y^1) = 0$, thus reject $y^2$ and move to next stage.
            \item if $\pi(x) > \pi(y^1) \geq \pi(y^0)$ then accept $y^1$ with probability
                                                            $ \alpha^1(x,y^0,y^1) =    \tfrac  {\pi(y^1)-\pi(y^0)}
                                                                                             {\pi(x)-\pi(y^0)}   $
        \end{enumerate}

        Therefore, Equation \eqref{accProbDR1Mat} can be also written as,
        \begin{equation}
            \alpha^1(x,y^0,y^1) = 1 \wedge    \dfrac  { 0 \vee [\pi(y^1) - \pi(y^0)] }     
                                                    { \pi(x) - \pi(y^0) }                
            \label{accProbDR1Mat2}
        \end{equation}

        Similarly, for any $i^{th}$ DR stage, we can generalize Equation \eqref{accProbDRi} as,
        \begin{equation}
            \alpha^1(x,y^0,y^1) = 1 \wedge    \dfrac  { 0 \vee [\pi(y^i) - \pi(y^*)] }     
                                                    { \pi(x) - \pi(y^*) }                
            \label{accProbDRiMat}
        \end{equation}
        \noindent where,
        \begin{equation}
            y^* = \argmax_{j<i}\pi(y_j)
        \end{equation}


        \begin{algorithm}
            \SetAlgoLined
            \caption{MCMC Algorithm for MatDRAM}
            \label{alg: MatDRAM}
            \textbf{Input\ \ \ :} \texttt{getLogFunc}: The objective function                       \   \   \textit{(mandatory)}    \\
            \textbf{Input\ \ \ :} $\texttt{ndim}$: Number of dimensions of the objective function   \   \   \textit{(mandatory)}    \\
            \textbf{Input\ \ \ :} $\chainSize$: Sample length                                       \   \   \textit{(optional)}     \\
            \textbf{Input\ \ \ :} $X_0$: Starting point                                             \   \   \textit{(optional)}     \\
            \textbf{Input\ \ \ :} $C_0$: Starting covariance matrix                                 \   \   \textit{(optional)}     \\
            \textbf{Input\ \ \ :} $N_{\mathrm{stages}}$: Number of DR stages                        \   \   \textit{(optional)}     \\
            \textbf{Input\ \ \ :} $\gamma(1:N_{\mathrm{stages}})$: DR Scale Factor Vector           \   \   \textit{(optional)}     \\
            \textbf{Input\ \ \ :} \texttt{AUP}: Adaptive update period                              \   \   \textit{(optional)}     \\
            \textbf{Input\ \ \ :} \texttt{randomSeed}                                               \   \   \textit{(optional, but needed for restart mode)}\\
            \textbf{Output:} $X_1$,..., $X_{N_{\mathrm{sample}}}$\\
            \bigbreak
            \textbf{\textit{Initialize:}}\\
            $q^0(.\mid x_{0})\leftarrow \mathscr{D}$ \ \ \textit{(where, $\mathscr{D}$ can be a normal proposal, $\mathcal{N}(X_0,C_0)$ or uniform proposal, $\mathcal{U}(X_0,C_0)$. )}\\
            \bigbreak
            \For{i $\leftarrow$ 1 to $N_{\mathrm{sample}}$} {
                Propose $Y^0_{\mathrm{cand}}\sim q^0(\cdot\mid x_{i-1})$\\
                Evaluate acceptance probability, $\alpha^0$\\
                \uIf{u $\sim \mathcal{U}(0,1) < \alpha^0$}{
                    Accept candidate, $X_i \leftarrow Y^0_{\mathrm{cand}}$\\
                    \textbf{break for}    \ \ \textit{(No need for any of the DR stages.)}
                }\Else{
                    \For{k = 1 to $N_{\mathrm{stage}}$} {
                        $q^k(\cdot\mid x_{i-1}) \leftarrow \gamma^k ~ q^{k-1}(\cdot\mid x_{i-1}) $\\
                        Propose $Y^k_{\mathrm{cand}} \sim q^k(\cdot\mid x_{i-1})$\\
                        Evaluate acceptance probability, $\alpha^k$\\
                        \uIf{u $\sim \mathcal{U}(0,1) < \alpha^k$}{
                            Accept candidate, $X_i \leftarrow Y^k_{\mathrm{cand}}$\\
                            \textbf{break for}    \ \ \textit{(No need for further DR stages.)}
                        }\ElseIf{k = $N_{\mathrm{stage}}$}{
                            Reject candidate, $X_i \leftarrow X_{i-1}$
                        }
                    }
                }
                \If{ ( i \text{mod \texttt{AUP}} ) = 0}{
                    Adapt proposal distribution $q^0$\\
                }
            }
        \end{algorithm}


%
%

    \subsection{The MatDRAM simulation restart functionality}

        A unique feature of the MatDRAM sampler is its restart functionality. If any runtime interruptions happen at any stage during a simulation, the sampler will be able to restart from where it left off and continue to generate the same chain that it would have generated had the simulation not been interrupted. Despite the inherently stochastic nature of Monte Carlo simulations, the MatDRAM algorithm has been implemented such that the resulting chain from a restarted simulation would be fully deterministic and identical to the chain of an uninterrupted simulation of the same configuration, up to 16 digits of precision. This restart functionality is identical to the restart functionality of the ParaDRAM algorithm of the ParaMonte library in C, C++, Fortran \citep{2020arXiv200809589S, 2020arXiv200914229S}.
        \newpar

        To restart a simulation, the user only needs to rerun the simulation by providing the same output file names. To generate a fully deterministic simulation, the user will have to also request the full double-precision accuracy for the simulation output files, set the seed of the random number generator and, enable the simulation's restart option prior to running the original simulation (before the interruption happens). After restarting the simulation, MatDRAM will automatically detect the previously-interrupted simulation output files. If all output files are detected, a message will be displayed indicating that the previous run is already completed and a new output file name will be asked. However, if certain files are missing, MatDRAM will attempt to restart the simulation from where it left off by collecting all of the previously-sampled states from the existing output chain file. The workflow of the restart functionality is given in Algorithm \ref{alg: restart}.
        \newpar

        MatDRAM collects the required restart information from the output restart file which is built continuously and dynamically throughout the uninterrupted simulation. To minimize the impacts of the restart IO on the performance and the external memory requirements of the algorithm, the restart file is automatically written in binary format. Alternatively, the user can request an \texttt{ASCII} restart file format (overriding the default binary format) in the input simulation specifications to the sampler. In such case, additional information about the dynamics of the proposal adaptations will be also written to the output restart file which can be later post-processed to understand the dynamic behavior of the MatDRAM algorithm.

        \begin{algorithm}
            \SetAlgoLined
            \caption{Restart Functionality}
            \label{alg: restart}
            \textbf{Input\ \ \ :} \texttt{getLogFunc}: The objective function                            \ \ \textit{(mandatory)}\\
            \textbf{Input\ \ \ :} \texttt{ndim}: Number of dimensions of the objective function\ \ \textit{(mandatory)}\\
            \textbf{Input\ \ \ :} \texttt{outputFileName}: Name of the output file                \ \ \textit{(mandatory)}\\
            \textbf{Input\ \ \ :} \texttt{randomSeed}                                            \ \ \textit{(mandatory)}\\

            \uIf{\texttt{outputFileName} exists}{
                Find the previously-sampled states in \texttt{outputFileName}, say $\chainSize^\mathrm{old}$ \\
                \uIf{$\chainSize^\mathrm{old}\neq0$}{
                    Copy previously sampled $\chainSize^\mathrm{old}$ states.\\
                    Continue with algorithm \ref{alg: MatDRAM} with $i=\chainSize^\mathrm{old}+1$)\\
                }\Else{
                    Run fresh run of MatDRAM via algorithm \ref{alg: MatDRAM}\\
                }
            }\Else{
                Run fresh run of MatDRAM via algorithm \ref{alg: MatDRAM}\\
            }
        \end{algorithm}

    \subsection{Efficient compact storage of the Markov Chain}

        One of the major design goals of the ParaMonte library and hence, MatDRAM, is the high performance of the library, i.e., the ability to handle large-scale simulations. Therefore, continuous external storage of the simulation output is essential for the ability of MatDRAM to handle large-scale simulations that exceed the random-access-memory (RAM) of the processor, as well as for the restart functionality. Such continuous external file input/ouput (IO) presents two major challenges:

        \begin{enumerate}[parsep=0ex,itemsep=1ex,topsep=1ex]
            \item
                Given the currently available computer technologies, external file IO is typically 2-4 orders of magnitude slower than the RAM storage. Therefore, the computational speed of the MatDRAM algorithm can significantly degrade for simulations involving complex high-dimensional objective functions.
            \item
                Moreover, the size of the resulting output Markov chain files can easily grow to several Gigabytes, making the storage of multiple simulation output files over the long term challenging or impossible.
        \end{enumerate}

        To overcome the above two challenges, we have introduced a very compact format to store the resulting Markov chain from MatDRAM simulations. This compact format significantly improves the library's performance and lowers the external storage requirements of the output files by 5-10 times without compromising the fully-deterministic restart functionality of MatDRAM or its ability to handle large-scale memory-demanding simulations.

        The idea behind the {\it compact} (as opposed to {\it verbose} or Markov) storage of the chain is to reduce the outputting of repeated redundant sample states in the Markov chain. The majority of the states in a typical Markov chain are identical because of the repeated rejections during the sampling. The fraction of repeated states in the verbose Markov chain is directly proportional to the rejection rate (or inversely to the acceptance probability). Therefore, we can recover an entire Markov chain by keeping track of only the uniquely accepted states. However, to ensure that the Markov chain can be later properly reconstructed, each uniquely sampled point is weighted by the number of times it is repeated in the actual verbose Markov chain. This approach both reduces the RAM and external memory requirements and improves the performance of the algorithm by reducing the frequency and the amount of IO.

        Nevertheless, for the sake of completeness, the MatDRAM also provides the option to specify a \texttt{verbose} format for the output chain files via the input specifications, in which case, the resulting Markov chain will be written to the output file {\it as is}. This \texttt{verbose} format of chain IO is not recommended except for debugging or exploration purposes since it significantly degrades the algorithm's performance and increases the memory requirements of the output files.

    \subsection{The final sample refinement}
        \label{sec:methods:refinement}

        A key element in Markov Chain Monte Carlo simulations is that the resulting chain is stochastic and has the Markovian property, which means that the future states depend only on the current state, not the previously visited states. Nevertheless, the resulting sample from a finite-size MCMC simulation is still highly autocorrelated, since each $X_n$ depends on its predecessor $X_{n-1}$. This dependence between all the successive pairs, ($X_n$, $X_{n-1}$), induces a significant non-zero correlation within the MCMC samples.
        \newpar

        For an infinite-length Markov chain that has converged to its stationary equilibrium distribution, the autocorrelation function is defined as,
        \begin{equation}
            \label{eq:randomSeqAC}
            \acf(k) = \frac{ \mathbb{E} \big[ (X_i - \mu) (X_{i+k} - \mu) \big] } { \sigma^2 } ~,
        \end{equation}

        \noindent where $(\mu,\sigma^2=\acf(0))$ represent the mean and the standard deviation of the Markov chain and $\mathbb{E}[\cdot]$ represents the expectation operator. The {\it Integrated Autocorrelation (IAC)} of the chain is defined with respect to the variance of the estimator of the mean value $\mu$,
        \begin{equation}
            \label{eq:randomSeqIAC1}
            \iac = 1 + 2\sum_{k=1}^{+\infty} \acf(k) ~,
        \end{equation}

        \noindent such that,
        \begin{equation}
            \label{eq:randomSeqIAC2}
            \lim_{n\rightarrow+\infty} \sqrt{\frac{n}{\iac}} \frac{\mu_n - \mu}{\sigma} \Rightarrow N(0,1) ~,
        \end{equation}

        \noindent where $\mu_n$ represents the sample mean of the chain of length $n$ and `$\Rightarrow$' stands for convergence in distribution.

        The value of IAC roughly indicates the number of Markov transitions (jumps) required to obtain an independent and identically distributed (i.i.d.) sample from the target distribution of the Markov chain. One wishes to obtain a finite MCMC sample whose size is at least of the order of the IAC of the chain \citep{robert2010introducing}. However, achieving this is often out of reach since the \iac~of the Markov chain is not known a prior. Alternatively, a more approachable goal would be to generate a chain with a predefined length and de-correlating it many times until a final {\it refined} i.i.d. sample from the target density is obtained.
        \newpar

        The numerical computation of \iac, however, comes with another challenge since the variance of its estimator diverges to infinity. Numerous techniques have been proposed to estimate the \iac~for the purpose of MCMC sample refinement. Among the most popular methods are the Batch Means (BM) \citep{fishman1978principles}, the Overlapping Batch Means (OBM) \citep{schmeiser1982batch}, the spectrum fit method \citep{heidelberger1981spectral}, the initial positive sequence estimator \citep{geyer1992practical}, as well as the auto-regressive processes \citep[e.g.,][]{plummer2006coda}.
        \newpar

        A series of comparison tests have been done by \citep{thompson2010comparison} to find out the fastest and most accurate method of estimating the \iac. Their results indicate that while the auto-regressive process appears to be the most accurate method of estimating the \iac, the Batch Means method provides a fair balance between the computational efficiency and numerical accuracy of the estimate.
        \newpar

        In MatDRAM, we have implemented the Batch Means method as the default method of estimating the \iac~of the resulting Markov chains based on the findings of \citep{thompson2010comparison}. Since almost all IAC estimation methods, including the Batch Means, generally underestimate its value, we have adopted a default aggressive methodology in MatDRAM algorithm where the autocorrelation of the chain is removed repeatedly until the final repeatedly-refined Markov chain does not exhibit any autocorrelation.
        \newpar

        The refinement of the chain is performed in two stages:
        \begin{enumerate}[parsep=0ex,itemsep=1ex,topsep=1ex]
            \item At the first stage, the {\it full Markov} chain is repeatedly refined based on the estimated \iac~values from the (non-Markovian) {\it compact chain} of the uniquely accepted points. This stage essentially removes any autocorrelation in the Markov chain that is due to the choice of too-small step sizes for the proposal distribution.

            \item Once the compact chain of accepted points is devoid of any autocorrelations, the second phase of the Markov chain refinement begins, with the \iac~values now being computed from the partially-refined ({\it verbose}) Markov chain, starting with the resulting refined Markov chain from the first stage of the refinement in the above.
        \end{enumerate}
        \newpar

        We have empirically found by numerous experimentations that the above approach often leads to final refined MCMC samples that are fully decorrelated while not being refined too much due to aggressive repetitive decorrelation of only the full Markov chain. An example comparison of the autocorrelation of the refined sample with that of the Markov chain will be given in \S\ref{sec:results}.

%% file: sec/results.tex
\section{Results}
\label{sec:results}

    We now describe the usage of MatDRAM via an example, including the initialization, the simulation specifications setup, the structure of the output files, and the post-processing of the results.

    \subsection{Calling the MatDRAM sampler}
    \label{routines}

        A MatDRAM sampler object can be readily instantiated by the following command,
        \begin{lstlisting}[]
pm = paramonte(); % generate an instance of the paramonte class
pmpd = pm.ParaDRAM(); % generate a MatDRAM sampler instance
        \end{lstlisting}

        The first command generates an instance of the \texttt{paramonte} class to which the MatDRAM sampler belongs. The second command instantiates a MatDRAM sampler object.
        By default, the simulation specifications are all automatically set to the appropriate default values. If needed, one can assign the simulation specifications as attributes of the \texttt{spec} component of the MatDRAM sampler object that was constructed in the above.
        \begin{lstlisting}[]
pmpd.spec.outputFileName = "./out/exampleRun";
pmpd.spec.chainSize = 10000;
        \end{lstlisting}

        There are currently a total of 36 optional MatDRAM simulation specifications that can be set manually by the user. The full set of specifications are extensively discussed on the documentation website of the MatDRAM library (as part of the ParaMonte library) at \url{https://www.cdslab.org/paramonte/notes/usage/paradram/specifications/}.

        Once the simulation specifications are set up, one can invoke the \mcode{runSampler()} method of the MatDRAM sampler object to initiate the DRAM-MCMC simulation. This method takes only two mandatory input arguments: \mcode{ndim} representing the number of dimensions of the objective function and \mcode{@getLogFunc} representing the handle to a MATLAB-implementation of the target objective function that is to be sampled. This function must take an input {\it column-vector} \texttt{point} of length \texttt{ndim} and return the natural logarithm of value of the objective function at the input point.
        \begin{lstlisting}[]
pmpd.runSampler(ndim, @getLogFunc);
        \end{lstlisting}

The most trivial example of such function is a MATLAB anonymous function representing a Multivariate Normal distribution of arbitrary dimensions. For example,
        \begin{lstlisting}[]
pm = paramonte();
pmpd = pm.ParaDRAM();
pmpd.runSampler(2, @(x)-sum(x.^2));
        \end{lstlisting}

        \noindent will sample a bivariate Normal density function. A more interesting yet simple example for illustration purposes would be a MultiVariate Normal (MVN) distribution whose  dimensions are correlated with each other. Let $X$ be an n-dimensional random vector with a MVN distribution (with mean $\mu$ and covariance $\Sigma$).
        \begin{equation}
            X \sim \mathcal{N}(\mu,\Sigma)
        \end{equation}
        where,
        \begin{itemize}[parsep = 0ex, itemsep = 0ex, topsep = 0ex]
            \item $X$ is an $n \times 1$ vector.
            \item $\mu$ is an $n \times 1$ vector, $E(X) = \mu$
            \item $\Sigma$ is an $n \times n$ matrix, $\Sigma = Cov(X)$.
        \end{itemize}

        Then the Probability Density Function (PDF) of $X$ is given by,
        \begin{equation}
            \label{pdf_mvn}
            f(x) =  \frac{1}{|\Sigma|^{\frac{1}{2}} (2 \pi)^{\frac{n}{2}}}                
                    \exp\left\{ -\frac{1}{2} (x-\mu)^\prime    \Sigma^{-1}(x-\mu)\right\}    
        \end{equation}

        An implementation of the above PDF for a \mcode{ndim}=4 dimensional MVN distribution with $\mu=$ \mcode{MEAN} and $\Sigma =$ \mcode{COV} is given below,

        \begin{lstlisting}[breakatwhitespace=false, numbers=none, label={lst:mvnlogfunc}]
function logFunc = getLogFunc(X)
    % This function returns the logarithm of the PDF(MVN)
    NDIM = 4;
    MEAN = [ 0.50, 0.00, -0.2, 0.30 ]';
    COV  = [ 1.00, 0.45, -0.3, 0.00 ...
           ; 0.45, 1.00, 0.30, -0.2 ...
           ; -.30, 0.30, 1.00, 0.60 ...
           ; 0.00, -.20, 0.60, 1.00 ...
           ];
    INVCOV = inv(COVMAT);
    % the log of the coefficient used in the MVN PDF
    MVN_COEF = NDIM*log(1./sqrt(2.*pi))+log(sqrt(det(INVCOV)));

    normedX = MEAN - X;
    logFunc = MVN_COEF - 0.5*(dot(normedX', INVCOV*normedX));
end
        \end{lstlisting}

\begin{figure}
            \centering
            \tcbox[colback = white, top=0pt,left=0pt,right=0pt,bottom=0pt]{\includegraphics[width=0.98\linewidth]{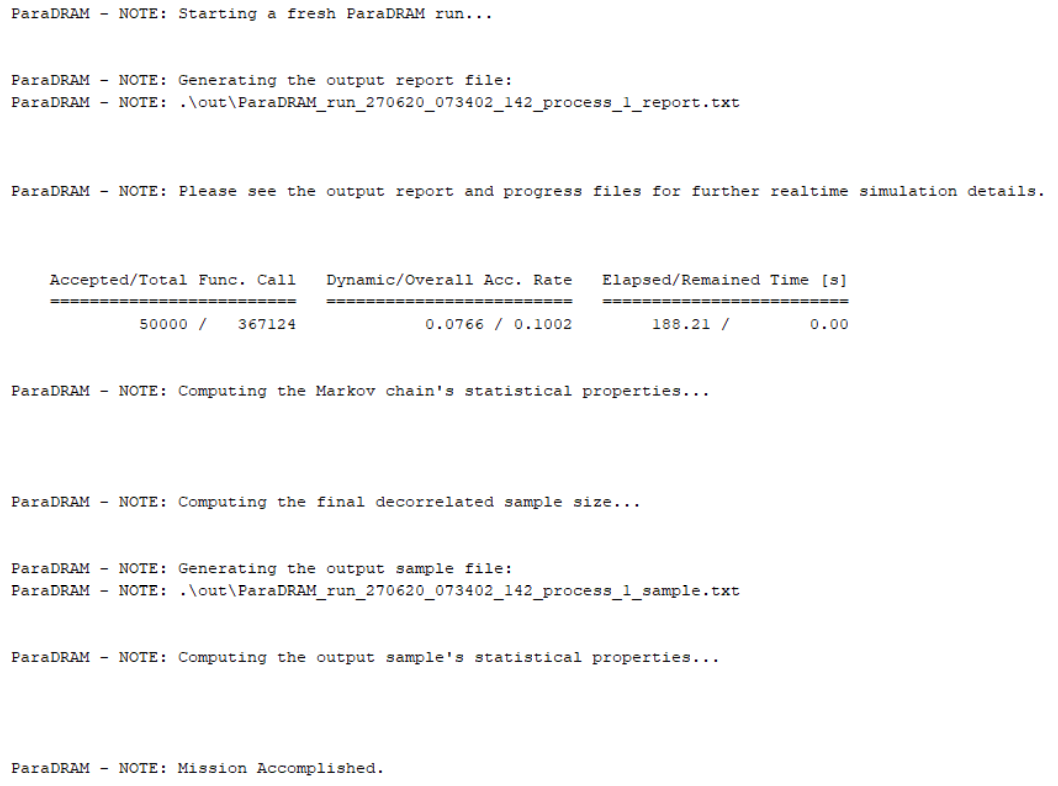}}
            \caption{An example MATLAB output screen after a call made to the pmpd.runSampler(ndim, @getLogFunc) method.}
            \label{fig:matlab_output}
        \end{figure}

        By saving this function in a file named as \texttt{getLogFunc.m} we can begin sampling it as follows,

        \begin{lstlisting}[]
pm = paramonte();
pmpd = pm.ParaDRAM();
ndim = 4;
pmpd.runSampler(ndim, @getLogFunc);
        \end{lstlisting}

        During the simulation, information about the progress made in sampling is dynamically displayed on the MATLAB command line. Figure \ref{fig:matlab_output} shows a snippet of the MATLAB output screen after a successful simulation run.

    \subsection{The MatDRAM simulation output files}

        Every successful MatDRAM simulation generates 5 output files, namely, the progress, report, sample, chain, and the restart files with the following suffixes respectively.
        \begin{enumerate}[parsep=0ex,itemsep=0ex,topsep=0ex]
            \item $\_progress.txt$
            \item $\_report.txt$
            \item $\_sample.txt$
            \item $\_chain.txt$
            \item $\_restart.txt$
        \end{enumerate}

        Each of the output files is prefixed with the file name that the user has provided. If no output filename has been specified by the user, an automatically-generated file name with a format similar to the following is used,
        \begin{lstlisting}[numbers=none]
MatDRAM_run_20201007_152848_916_process_1_progress.txt
MatDRAM_run_20201007_152848_916_process_1_restart.txt
MatDRAM_run_20201007_152848_916_process_1_report.txt
MatDRAM_run_20201007_152848_916_process_1_sample.txt
MatDRAM_run_20201007_152848_916_process_1_chain.txt
        \end{lstlisting}

        \begin{enumerate}[parsep=0ex,itemsep=1ex,topsep=0ex]

            \item
                The output report file:\\
                This file contains all the details about the simulation setup, in the following order,
                \begin{itemize}[parsep=0ex,itemsep=1ex,topsep=0ex]
                    \item The MatDRAM banner and version specifications,
                    \item The specifications of the processor on which the current simulation is being performed,
                    \item The specifications of the current MatDRAM simulation being performed along with their values and descriptions,
                    \item The relevant details about the simulation timing and performance, if the simulation finishes successfully, along with,
                    \item The statistics of the simulation results, and,
                    \item The final message: “Mission Accomplished.” indicating the successful ending of the simulation.
                \end{itemize}

            \item
                The output sample file:\\
                This is the primary output file of interest produced by the MatDRAM sampler as it contains a refined, decorrelated, independent and identically-distributed (i.i.d) set of random states (points) from the user-provided mathematical objective function. This file contains only two pieces of information,
                \begin{itemize}[parsep=0ex,itemsep=0ex,topsep=0ex]
                    \item
                        \texttt{SampleLogFunc} – A data column representing the values of the user-provided mathematical objective function at the corresponding sampled states on each row of the file,
                    \item
                        The sampled state – a row-wise vector of values that represents the current state that has been sampled corresponding to each \texttt{SampleLogFunc} on each row of the file.
                \end{itemize}

            \item
                The output progress file:\\
                This file contains realtime information about the progress of the simulation, including,
                \begin{itemize}[parsep=0ex,itemsep=0ex,topsep=0ex]
                    \item information about the number of calls the MatDRAM sampler makes to the user-provided mathematical objective function,
                    \item information about the overall efficiency of the MatDRAM sampler,
                    \item information about the dynamic efficiency of the sampler over the past \texttt{progressReportPeriod} number of calls to the mathematical objective function,
                    \item information about the timing of the simulation including,
                    \begin{itemize}[parsep=0ex,itemsep=0ex,topsep=0ex]
                        \item the time spent since the start of the simulation,
                        \item the time since the last progress report,
                        \item the estimated time to finish the simulation.
                    \end{itemize}
                \end{itemize}

            \item
                The output restart file:\\
                The output restart file contains all information that is needed to restart a simulation should a runtime interruption happen. When the restart file format is set to \texttt{ascii}, the MatDRAM sampler also writes additional information about the dynamics of the proposal distribution updates to the restart file. This information is not required for the simulation restart, however, it can be used in the postprocessing phase of the simulation to better understand the inner-workings of the DRAM sampler and visualize the dynamics of the proposal distribution updates.

            \item
                The output chain file:\\
                The output chain file contains information about all accepted states that have been visited by the MatDRAM sampler, with the following columns of data,
                \begin{itemize}[parsep=0ex,itemsep=0ex,topsep=0ex]
                    \item \texttt{ProcessID} – the ID of the processor that has successfully sampled the current state (point) from the user-provided mathematical objective function. In the current implementation of MatDRAM, which is serial, this ID is always 1. However it is kept in this output file for purpose of consistency with the ParaMonte \texttt{ParaDRAM} sampler and the possibility of expanding the MatDRAM library to a parallel version without interrupting the structure of the input file.
                    \item \texttt{DelayedRejectionStage} – the delayed-rejection stage at which the newly sampled state has been accepted,
                    \item \texttt{MeanAcceptanceRate} – the mean acceptance rate of the sampler up to the newly-sampled state at a given row,
                    \item \texttt{AdaptationMeasure} – the amount of adaptation performed on the sampler’s proposal distribution, which is a number between zero and one, with one indicating extreme adaptation being performed at that stage in the simulation on the proposal distribution and, a value of zero indicating absolutely no adaptation being performed since the last sampled state,
                    \item \texttt{BurninLocation} – the runtime estimate of the number of sampled states (from the beginning of the simulation) that are potentially non-useful and must be discarded as the burnin period,
                    \item \texttt{SampleWeight} – the number of times each newly-sampled point is repeated in the Markov chain before the next candidate state is accepted,
                    \item \texttt{SampleLogFunc} – the value of the user-provided mathematical objective function at the currently-sampled state,
                    \item followed by a row-wise vector of values that represent the current state that has been sampled.
                \end{itemize}

        \end{enumerate}

    \subsection{The MatDRAM visualization tools}

        The MatDRAM sampler automatically ships with a number of visualization tools that enable seamless creation of sophisticated plots of the output results from a MatDRAM simulation. These include, histogram, trace-plot, line/scatter-plot in 2D or 3D, kernel density plot in 2D or 3D, grid plot, autocorrelation plot, as well as correlation and covariance matrix plots. A few of these visualization tools will be discussed in this section.

        \subsubsection{Grid Plot}

\begin{figure}
                \centering
                \tcbox[colback = white, top=0pt,left=0pt,right=0pt,bottom=0pt]{\includegraphics[width=0.98\linewidth]{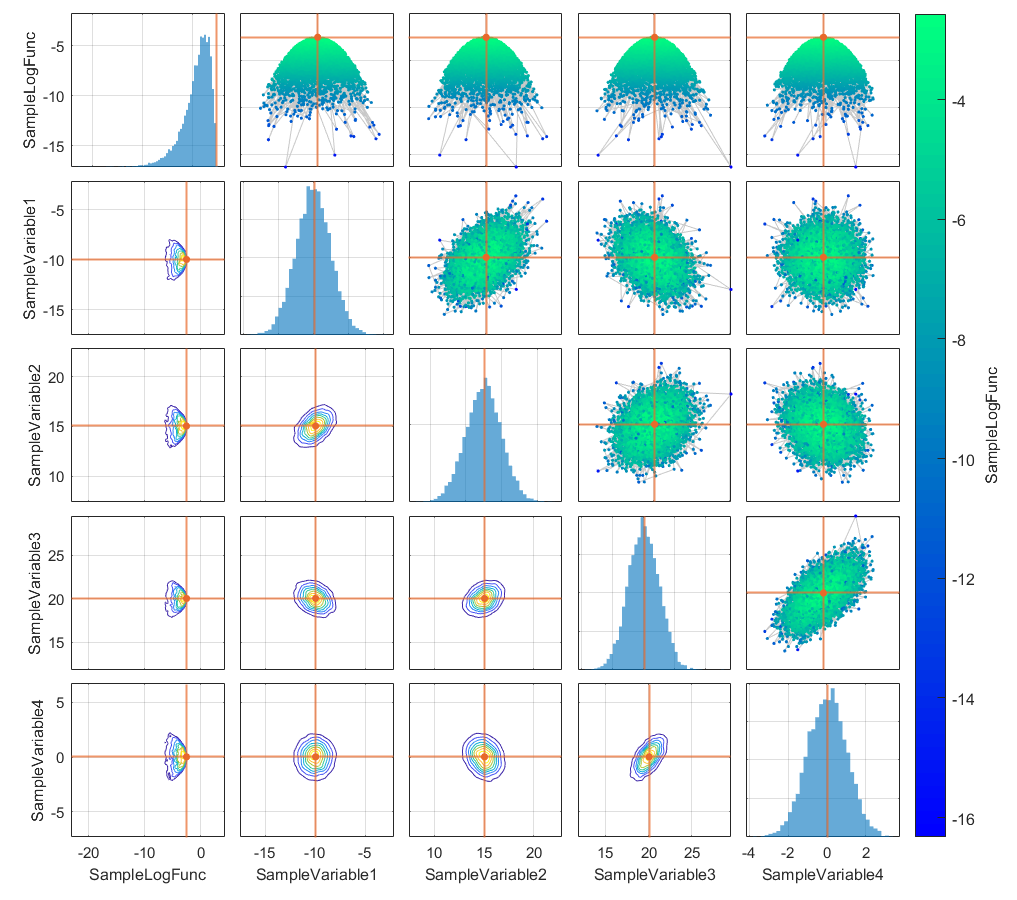}}
                \caption{An example of a Grid plot for the problem of sampling a 4-dimensional multivariate normal distribution.}
                \label{fig:plot_grid_mvn_sample}
            \end{figure}

            Figure \ref{fig:plot_grid_mvn_sample} shows an example of several types of grid plots that can be made from the output of MatDRAM simulations, in this particular case, a MCMC simulation of a 4-dimensional MVN distribution with the covariance matrix given in the code snippet \ref{lst:mvnlogfunc}. On each subplot, the mode of the distribution of sampled states is also represented by the orange lines and the single orange scatter points. The entire MatDRAM code to generate this plot from the output of the simulation is the following,

            \begin{lstlisting}[]
sample = pmpd.readSample(); % read the output sample
sample = sample{1}; % use the contents of the first output file found
sample.plot.grid.make(); % make a grid plot of all output variables
sample.plot.grid.addTarget(); % add target (the mode)
            \end{lstlisting}

\begin{figure}
                \centering
                \tcbox[colback = white, top=0pt,left=0pt,right=0pt,bottom=0pt]{\includegraphics[width = 0.98\linewidth]{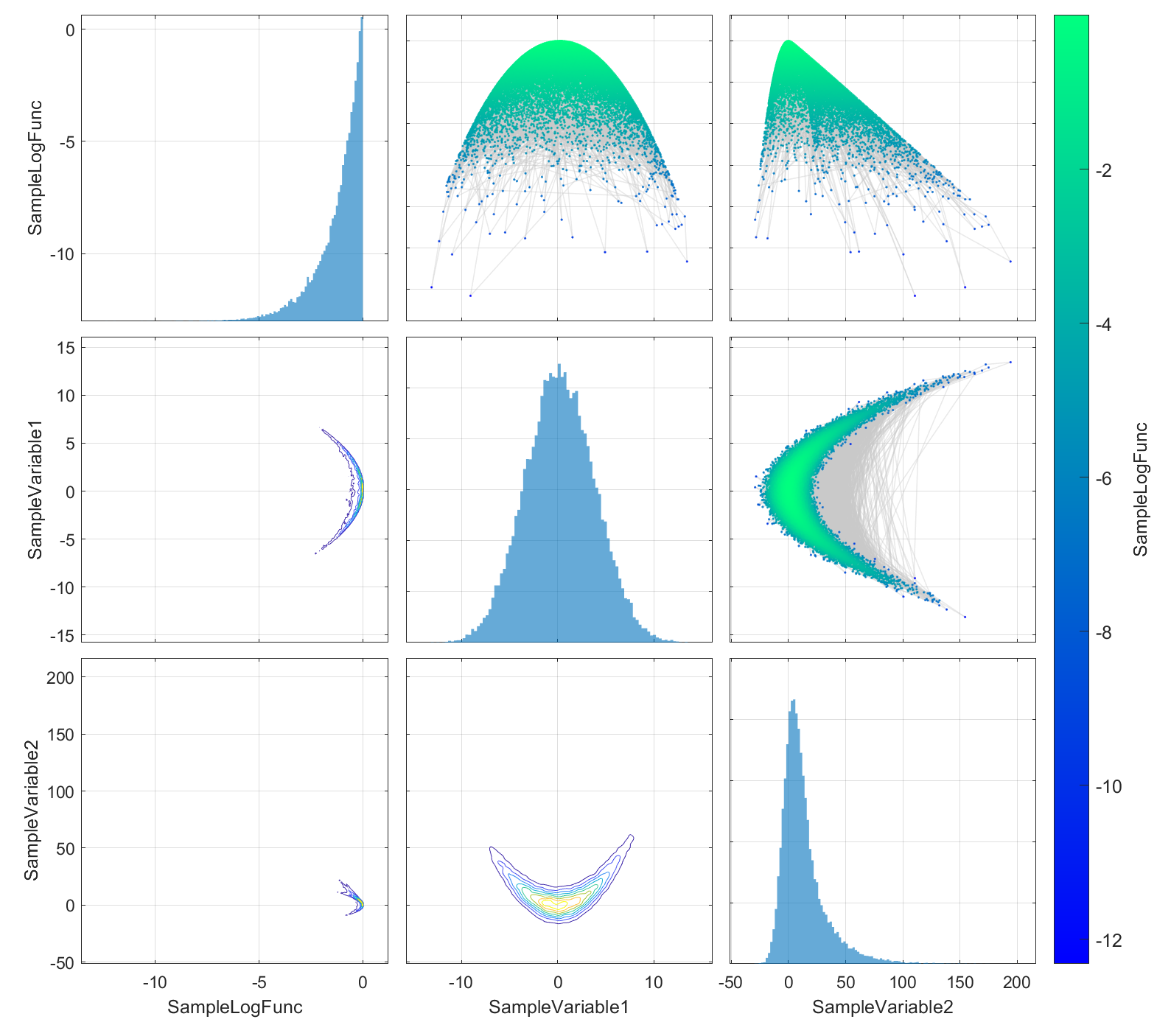}}
                \caption{An illustration of a Grid plot for output of the MCMC sampling of the Rosenbrock function in two dimensions.}
                \label{fig:plot_grid_rosen_sample}
            \end{figure}

            Figure \ref{fig:plot_grid_rosen_sample} displays another example of a grid plot for the problem of sampling the Rosenbrock function in 2-dimensions given by the following equation.
            \begin{equation}
                f(x) = c[(a-x)^2 + b(y-x^2)^2]
            \end{equation}

            \noindent where,
            \begin{equation}
                \left.
                    \begin{array}{ll}
                        a = 0.2 \\
                        b = 0.2 \\
                        c = -\frac{1}{20}
                    \end{array}
                \right \}
            \end{equation}

        \subsubsection{The Autocorrelation Plot}

            The goal of the MatDRAM MCMC sampler is to generate a refined sample that has no auto-correlation. As mentioned before, the output Markov chains of MCMC samplers are generally highly autocorrelated. It is, therefore, important to ensure the final refined sample is independent and identically distributed (i.i.d.). As discussed in \$\ref{sec:methods:refinement}, the MatDRAM library automatically, aggressively, and recursively refines the output Markov chains to generate the final refined i.i.d. sample. Figure \ref{fig:plot_ACF_line} compares the auto-correlation plots for the verbose (Markov) chain and the final refined sample for the example problem of sampling a 4-dimensional MVN distribution.

            \begin{figure}
                \centering
                \tcbox[colback = white, top=0pt,left=0pt,right=0pt,bottom=0pt]{
                    \begin{subfigure}{.49\linewidth}
                        \includegraphics[width=.98\linewidth]{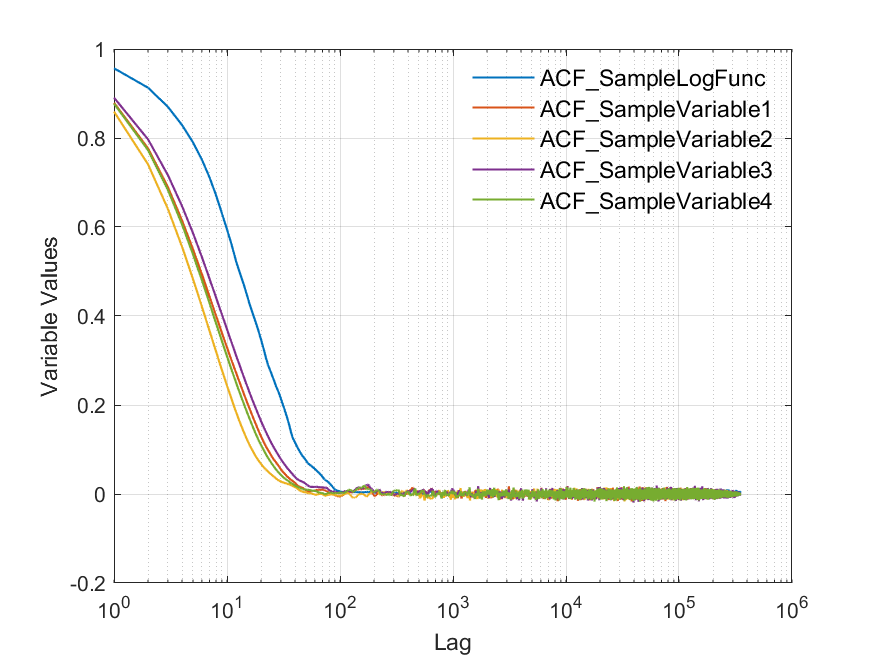}
                        \caption{The Autocorrelation of the MCMC chain.}
                        \label{fig:plot_ACF_line_MCMC}
                    \end{subfigure}
                    \hfill
                    \begin{subfigure}{.49\linewidth}
                        \includegraphics[width=.98\linewidth]{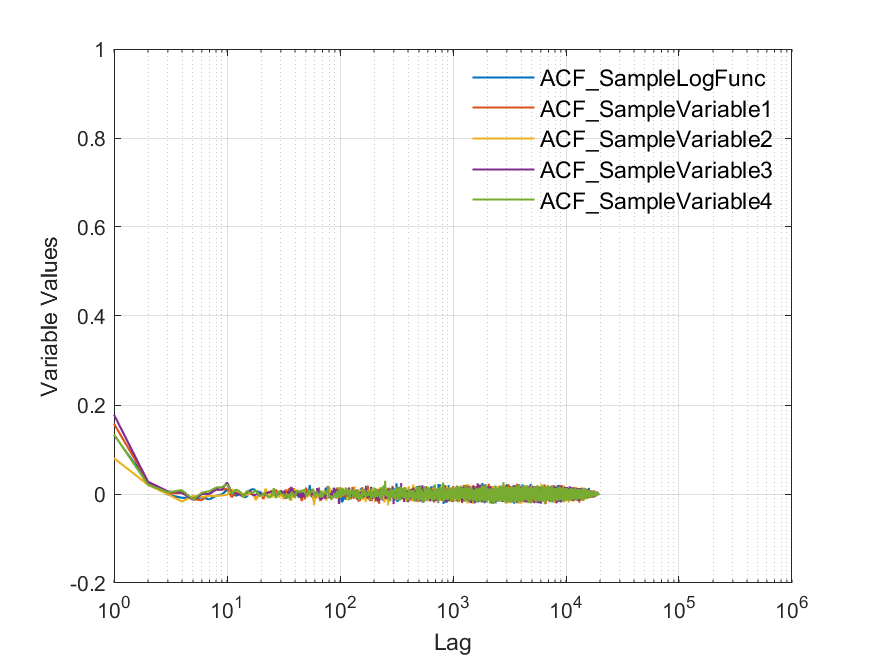}
                        \caption{The Autocorrelation of the refined sample.}
                        \label{fig:plot_ACF_line_sample}
                    \end{subfigure}
                }
                \caption{A comparison of the autocorrelations of the full output Markov chain and the final refined sample generated by the MatDRAM sampler for the example problem of sampling a 4-dimensional MVN distribution.}
                \label{fig:plot_ACF_line}
            \end{figure}

        \subsubsection{Monitoring the diminishing adaptation of the proposal distribution of the MatDRAM sampler}

            An important criterion for the ergodicity and reversibility of adaptive Markov chains, in particular, the MatDRAM algorithm presented here, is the diminishing adaptation of the proposal distribution of the sampler. Ideally, such proposal adaptivity should be measured via the {\it total variation distance} (TVD) between any two adjacent adaptively updated proposal distributions. The computational intractability of TVD, however, presents a major barrier toward practical implementation of this measure of adaptation.

            Here, we follow the approach of \cite{2020arXiv200809589S, shahmoradi2020paradram} to implement a computationally feasible upper limit on the value of TVD between any two subsequently updated proposal distributions within a MatDRAM simulation. This novel approach is based on the definition of the Hellinger distance \citep{hellinger1909neue}. Unlike TVD, the Hellinger distance has closed form expression, in particular, in the case of MultiVariate Normal distribution which is the most popular choice of proposal distribution for MCMC samplers, and the default proposal kernel of MatDRAM.

            For any two adjacent, adaptively-updated $d$-dimensional proposal distributions, $(\Pam_i,\Pam_{i+1})$, the Hellinger distance is defined by the following equation,
            \begin{equation}
                \label{eq:hellingerDefinition}
                {\mathrm{H}}^2 (\Pam_i,\Pam_{i+1}) = 1 - \int_{\mathbb{R}^{d}} \sqrt{ \pam_i(x)~ \pam_{i+1}(x) } ~ \diff x ~.
            \end{equation}

            \noindent where $\pam$ represents the corresponding probability density function. By definition, a value of zero for the Hellinger distance indicates the identity of the two probability distributions, whereas a value of one indicates the complete dissimilarity. A closer look at the above equation reveals that the Hellinger measure is in fact, the $L^2$ distance between the two probability distributions. The connection between the TVD and the Hellinger distance is made via the following inequality relationship \citep[e.g.,][]{2020arXiv200809589S},

            \begin{equation}
                \label{eq:tvdUpperBound}
                {\mathrm{TVD}}(\Pam_i,\Pam_{i+1}) \leq
                {\mathrm{H}}(\Pam_i,\Pam_{i+1}) \sqrt{1 - \frac{{\mathrm{H}}^2(\Pam_i,\Pam_{i+1})}{4} } ~.
            \end{equation}

\begin{figure}
                \centering
                \tcbox[colback = white, top=0pt,left=0pt,right=0pt,bottom=0pt]{
                    \begin{subfigure}{.49\linewidth}
                        \includegraphics[width=.98\linewidth]{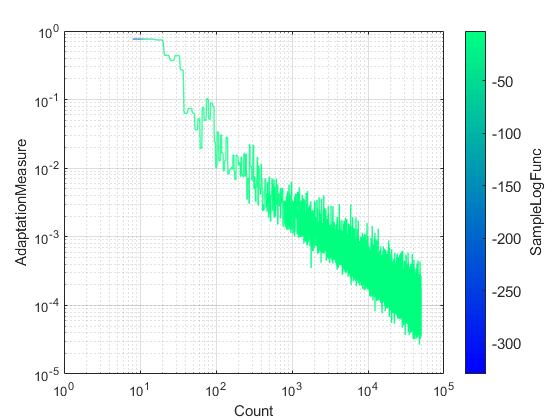}
                        \caption{The adaptation measure of proposal distribution.}
                        \label{fig:adaptationMeasure}
                    \end{subfigure}
                    \hfill
                    \begin{subfigure}{.49\linewidth}
                        \includegraphics[width=.98\linewidth]{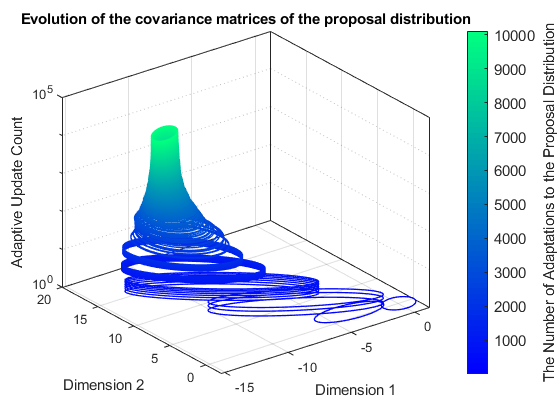}
                        \caption{The dynamic evolution of the covariance matrix of the Gaussian proposal distribution of the MatDRAM sampler, along the first two dimensions.}
                        \label{fig:covmatEvol}
                    \end{subfigure}
                }
                \caption{An illustration of the dynamic adaptation of the proposal distribution of the MatDRAM sampler.}
                \label{fig:plot_ACF_line}
            \end{figure}

            The computation of the above upper bound can be efficiently done in the case of MVN proposal distributions. This upper bound on TVD is automatically computed and reported to the output chain files of all MatDRAM simulations. Figure \ref{fig:adaptationMeasure}, depicts the evolution of the adaptation measure for the same problem of sampling a 4-dimensional MVN as before. The diminishing adaptation illustrated in this figure is an excellent example and indicator of successful adaptive MCMC sampling, where the adaptation upper bound is continuously and progressively decaying with the simulation progress.

            An example visualization of the dynamics of covariance matrices of the proposal distribution of the sampler for the same MVN sampling problem is provided in Figure \ref{fig:covmatEvol}.

            The adaptation measure upper bound that we have defined in the above can also be a strong indicator of the lack of convergence of the Markov chain to the target density, since in such cases, the amount of adaptation of the proposal distribution generally does not diminish and remains constant or wildly fluctuates with the simulation progress.

%% file: sec/discussion.tex
\section{discussion}
\label{sec:discussion}

    In this work we presented the algorithmic and implementation details of MatDRAM, an open-source MIT-licensed pure-MATLAB Monte Carlo simulation library that implements a variant of the Delayed-Rejection Adaptive Metropolis Markov Chain Monte Carlo algorithm. In developing the MatDRAM library, we have been careful to follow, as much as possible, the design goals of the ParaDRAM reference algorithm of the parent library of MatDRAM: the ParaMonte library. The toolbox and the postprocessing tools are all developed as part of the ParaMonte library, which is accessible to the public at \url{https://github.com/cdslaborg/paramonte}. Due to the intricate dependencies between MatDRAM and its parent ParaMonte library, both projects are currently maintained as a unified Git repository, although contributions to each project can be done separately and independently of the other. The procedure to build the library via the ParaMonte build script is simple. On Windows operating systems, the build process can be initiated by typing the following command on a Batch command-prompt,

    \begin{lstlisting}[frame=false]
install.bat --matdram
    \end{lstlisting}

    Similarly, the library can be built for linux/macOS platforms by typing the following command in a Bash terminal,

    \begin{lstlisting}[frame=false]
./install.sh --matdram
    \end{lstlisting}

    The latest build of MatDRAM is and will be available on \href{https://www.mathworks.com/matlabcentral/fileexchange/80866-matdram-delayed-rejection-adaptive-metropolis-mcmc}{the MATLAB FileExchange repository}. The public classes and functions of MatDRAM are fully documented and can be accessed via the traditional \texttt{help} or \texttt{doc} feature of MATLAB. The sampler as well as the simulation specifications and the structure of the output files are documented at \url{https://www.cdslab.org/paramonte/}.

    The MatDRAM visualization and postprocessing tools are capable of analyzing and plotting the Monte Carlo simulation results of not only the MatDRAM sampler, but also the reference ParaDRAM sampler \citep{2020arXiv200809589S, shahmoradi2020paradram} that is currently available in other programming languages, including C/C++/Fortran \citep{2020arXiv200914229S} and Python \citep{2020arXiv201000724S}.

    In designing the MatDRAM library, we have made sure to implement a fully-deterministic restart functionality for all MCMC simulations. This is an important feature for long computationally-expensive Monte Carlo simulations, in particular, on supercomputers where the allocated time is often limited to less than 24 hours. In the case of an unwanted simulation interruption, MatDRAM is able to seamlessly restart the incomplete simulation and produce the same chain that it would have generated, had the interruption not happened. This identity of the original and the restart simulation results is valid up to 16 digits of precision.

    While the MatDRAM algorithm has been developed to automate adaptive MCMC simulations as much as possible, one can always find instances of complex unusual objective functions that require special care and perhaps hand-tuning of the settings of the sampler. Areas of future work include the development of a parallel version of MatDRAM and further automation of adaptive MCMC simulations by automatically adjusting the scale of the proposal distributions of the sampler, in addition to the adaptation of the shape of the proposal distribution which is currently implemented in the MatDRAM algorithm.

%% file: main.bbl
\begin{thebibliography}{10}
\expandafter\ifx\csname url\endcsname\relax
  \def\url#1{\texttt{#1}}\fi
\expandafter\ifx\csname urlprefix\endcsname\relax\def\urlprefix{URL }\fi
\expandafter\ifx\csname href\endcsname\relax
  \def\href#1#2{#2} \def\path#1{#1}\fi

\bibitem{shahmoradi2017multilevel}
A.~Shahmoradi, Multilevel bayesian parameter estimation in the presence of
  model inadequacy and data uncertainty, arXiv preprint arXiv:1711.10599.

\bibitem{2017arXiv171110599S}
A.~{Shahmoradi}, {Multilevel Bayesian Parameter Estimation in the Presence of
  Model Inadequacy and Data Uncertainty}, arXiv e-prints (2017)
  arXiv:1711.10599\href {http://arxiv.org/abs/1711.10599}
  {\path{arXiv:1711.10599}}.

\bibitem{2020arXiv200809589S}
A.~{Shahmoradi}, F.~{Bagheri}, {ParaDRAM: A Cross-Language Toolbox for Parallel
  High-Performance Delayed-Rejection Adaptive Metropolis Markov Chain Monte
  Carlo Simulations}, arXiv e-prints (2020) arXiv:2008.09589\href
  {http://arxiv.org/abs/2008.09589} {\path{arXiv:2008.09589}}.

\bibitem{shahmoradi2020paradram}
A.~Shahmoradi, F.~Bagheri, Paradram: A cross-language toolbox for parallel
  high-performance delayed-rejection adaptive metropolis markov chain monte
  carlo simulations, arXiv preprint arXiv:2008.09589.

\bibitem{metropolis1949monte}
N.~Metropolis, S.~Ulam, The monte carlo method, Journal of the American
  statistical association 44~(247) (1949) 335--341.

\bibitem{von195113}
J.~Von~Neumann, 13. various techniques used in connection with random digits.

\bibitem{segre1955fermi}
E.~Segre, Fermi and neutron physics, Reviews of Modern Physics 27~(3) (1955)
  257.

\bibitem{metropolis1987beginning}
N.~Metropolis, The beginning of the monte carlo method.

\bibitem{metropolis1953equation}
N.~Metropolis, A.~W. Rosenbluth, M.~N. Rosenbluth, A.~H. Teller, E.~Teller,
  Equation of state calculations by fast computing machines, The journal of
  chemical physics 21~(6) (1953) 1087--1092.

\bibitem{hastings1970monte}
W.~K. Hastings, Monte carlo sampling methods using markov chains and their
  applications, Biometrika 57~(1) (1970) 97--109.

\bibitem{shahmoradi2013multivariate}
A.~Shahmoradi, A multivariate fit luminosity function and world model for long
  gamma-ray bursts, The Astrophysical Journal 766~(2) (2013) 111.

\bibitem{shahmoradi2015short}
A.~Shahmoradi, R.~J. Nemiroff, Short versus long gamma-ray bursts: a
  comprehensive study of energetics and prompt gamma-ray correlations, Monthly
  Notices of the Royal Astronomical Society 451~(1) (2015) 126--143.

\bibitem{2020arXiv200601157O}
J.~A. {Osborne}, A.~{Shahmoradi}, R.~J. {Nemiroff}, {A Multilevel Empirical
  Bayesian Approach to Estimating the Unknown Redshifts of 1366 BATSE Catalog
  Long-Duration Gamma-Ray Bursts}, arXiv e-prints (2020) arXiv:2006.01157\href
  {http://arxiv.org/abs/2006.01157} {\path{arXiv:2006.01157}}.

\bibitem{2019arXiv190306989S}
A.~{Shahmoradi}, R.~J. {Nemiroff}, {A Catalog of Redshift Estimates for 1366
  BATSE Long-Duration Gamma-Ray Bursts: Evidence for Strong Selection Effects
  on the Phenomenological Prompt Gamma-Ray Correlations}, arXiv e-prints (2019)
  arXiv:1903.06989\href {http://arxiv.org/abs/1903.06989}
  {\path{arXiv:1903.06989}}.

\bibitem{shahmoradi2014predicting}
A.~Shahmoradi, D.~K. Sydykova, S.~J. Spielman, E.~L. Jackson, E.~T. Dawson,
  A.~G. Meyer, C.~O. Wilke, Predicting evolutionary site variability from
  structure in viral proteins: buriedness, packing, flexibility, and design,
  Journal of molecular evolution 79~(3-4) (2014) 130--142.

\bibitem{lima2017ices}
E.~Lima, J.~Oden, B.~Wohlmuth, A.~Shahmoradi, D.~Hormuth~II, T.~Yankeelov, Ices
  report 17-14.

\bibitem{lima2017selection}
E.~Lima, J.~Oden, B.~Wohlmuth, A.~Shahmoradi, D.~Hormuth, T.~Yankeelov,
  L.~Scarabosio, T.~Horger, Selection and validation of predictive models of
  radiation effects on tumor growth based on noninvasive imaging data, Computer
  Methods in Applied Mechanics and Engineering.

\bibitem{haario2006dram}
H.~Haario, M.~Laine, A.~Mira, E.~Saksman, Dram: efficient adaptive mcmc,
  Statistics and computing 16~(4) (2006) 339--354.

\bibitem{soetaert2010inverse}
K.~Soetaert, T.~Petzoldt, et~al., Inverse modelling, sensitivity and monte
  carlo analysis in r using package fme, Journal of Statistical Software 33~(3)
  (2010) 1--28.

\bibitem{patil2010pymc}
A.~Patil, D.~Huard, C.~J. Fonnesbeck, Pymc: Bayesian stochastic modelling in
  python, Journal of statistical software 35~(4) (2010) 1.

\bibitem{miles2019pymcmcstat}
P.~R. Miles, pymcmcstat: A python package for bayesian inference using delayed
  rejection adaptive metropolis, Journal of Open Source Software 4~(38) (2019)
  1417.

\bibitem{stapor2018pesto}
P.~Stapor, D.~Weindl, B.~Ballnus, S.~Hug, C.~Loos, A.~Fiedler, S.~Krause,
  S.~Hro{\ss}, F.~Fr{\"o}hlich, J.~Hasenauer, Pesto: parameter estimation
  toolbox, Bioinformatics 34~(4) (2018) 705--707.

\bibitem{queso}
E.~Prudencio, K.~Schulz, The parallel {C}++ statistical library queso:
  Quantification of uncertainty for estimation, simulation and optimization,
  in: M.~Alexander, P.~D'Ambra, A.~Belloum, G.~Bosilca, M.~Cannataro,
  M.~Danelutto, B.~Martino, M.~Gerndt, E.~Jeannot, R.~Namyst, J.~Roman,
  S.~Scott, J.~Traff, G.~Vallée, J.~Weidendorfer (Eds.), Euro-Par 2011:
  Parallel Processing Workshops, Vol. 7155 of Lecture Notes in Computer
  Science, Springer Berlin Heidelberg, 2012, pp. 398--407.

\bibitem{shahmoradi2019paramonte}
A.~Shahmoradi, Paramonte: A user-friendly parallel monte carlo optimization,
  sampling, and integration library for scientific inference, Bulletin of the
  American Physical Society.

\bibitem{kumbhare2020parallel}
S.~Kumbhare, A.~Shahmoradi, Parallel adapative monte carlo optimization,
  sampling, and integration in c/c++, fortran, matlab, and python, Bulletin of
  the American Physical Society.

\bibitem{shahmoradi2020paramonte}
A.~Shahmoradi, F.~Bagheri, S.~Kumbhare, Paramonte: Plain powerful parallel
  monte carlo library, Bulletin of the American Physical Society.

\bibitem{2020arXiv200914229S}
A.~{Shahmoradi}, F.~{Bagheri}, {ParaMonte: A high-performance serial/parallel
  Monte Carlo simulation library for C, C++, Fortran}, arXiv e-prints (2020)
  arXiv:2009.14229\href {http://arxiv.org/abs/2009.14229}
  {\path{arXiv:2009.14229}}.

\bibitem{shahmoradi2020paramonteII}
A.~Shahmoradi, F.~Bagheri, Paramonte: A high-performance serial/parallel monte
  carlo simulation library for c, c++, fortran, arXiv preprint
  arXiv:2009.14229.

\bibitem{2020arXiv201000724S}
A.~{Shahmoradi}, F.~{Bagheri}, J.~A.~e. {Osborne}, {Fast fully-reproducible
  serial/parallel Monte Carlo and MCMC simulations and visualizations via
  ParaMonte::Python library}, arXiv e-prints (2020) arXiv:2010.00724\href
  {http://arxiv.org/abs/2010.00724} {\path{arXiv:2010.00724}}.

\bibitem{haario1999adaptive}
H.~Haario, E.~Saksman, J.~Tamminen, Adaptive proposal distribution for random
  walk metropolis algorithm, Computational Statistics 14~(3) (1999) 375--396.

\bibitem{haario2001adaptive}
H.~Haario, E.~Saksman, J.~Tamminen, et~al., An adaptive metropolis algorithm,
  Bernoulli 7~(2) (2001) 223--242.

\bibitem{tierney1999some}
L.~Tierney, A.~Mira, Some adaptive monte carlo methods for bayesian inference,
  Statistics in medicine 18~(17-18) (1999) 2507--2515.

\bibitem{green2001delayed}
P.~J. Green, A.~Mira, Delayed rejection in reversible jump
  metropolis--hastings, Biometrika 88~(4) (2001) 1035--1053.

\bibitem{mira2001metropolis}
A.~Mira, et~al., On metropolis-hastings algorithms with delayed rejection,
  Metron 59~(3-4) (2001) 231--241.

\bibitem{gelman1996efficient}
A.~Gelman, G.~O. Roberts, W.~R. Gilks, et~al., Efficient metropolis jumping
  rules, Bayesian statistics 5~(599-608) (1996) 42.

\bibitem{peskun1973optimum}
P.~H. Peskun, Optimum monte-carlo sampling using markov chains, Biometrika
  60~(3) (1973) 607--612.

\bibitem{robert2010introducing}
C.~P. Robert, G.~Casella, G.~Casella, Introducing monte carlo methods with r,
  Vol.~18, Springer, 2010.

\bibitem{fishman1978principles}
G.~S. Fishman, Principles of discrete event simulation.[book review].

\bibitem{schmeiser1982batch}
B.~Schmeiser, Batch size effects in the analysis of simulation output,
  Operations Research 30~(3) (1982) 556--568.

\bibitem{heidelberger1981spectral}
P.~Heidelberger, P.~D. Welch, A spectral method for confidence interval
  generation and run length control in simulations, Communications of the ACM
  24~(4) (1981) 233--245.

\bibitem{geyer1992practical}
C.~J. Geyer, Practical markov chain monte carlo, Statistical science (1992)
  473--483.

\bibitem{plummer2006coda}
M.~Plummer, N.~Best, K.~Cowles, K.~Vines, Coda: convergence diagnosis and
  output analysis for mcmc, R news 6~(1) (2006) 7--11.

\bibitem{thompson2010comparison}
M.~B. Thompson, A comparison of methods for computing autocorrelation time,
  arXiv preprint arXiv:1011.0175.

\bibitem{hellinger1909neue}
E.~Hellinger, Neue begr{\"u}ndung der theorie quadratischer formen von
  unendlichvielen ver{\"a}nderlichen., Journal f{\"u}r die reine und angewandte
  Mathematik (Crelles Journal) 1909~(136) (1909) 210--271.

\end{thebibliography}
